\documentclass{icrc}

\usepackage{times}
\usepackage{graphicx} 
\newcommand{\psim}{\lower.5ex\hbox{$\; \buildrel \propto \over\sim \;$}}
\newcommand{\lesssim}{\lower.5ex\hbox{$\; \buildrel < \over\sim \;$}}
\newcommand{\gtrsim}{\lower.5ex\hbox{$\; \buildrel > \over\sim \;$}}
\newcommand{\e}{\epsilon}            
\newcommand{\ep}{\epsilon^\prime}

\begin{document}

\title{High-Energy Neutrino Production through Photopion Processes in Blazars}
\author[1]{C. D. Dermer}
\affil[1]{Code 7653, Naval Research Laboratory, Washington, DC 20375-5352 USA}
\author[2]{A. Atoyan}
\affil[2]{McGill University, 3600 University Str., Montreal H3A 2T8, Canada}

\correspondence{dermer@gamma.nrl.navy.mil}

\firstpage{1}
\pubyear{2001}


\maketitle

\begin{abstract}

The measured spectral energy distribution and variability time scale
are used to determine the radiation and magnetic-field energy
densities in the relativistic plasma that forms the gamma-ray emitting
jet in the blazar 3C 279. Assuming that protons are accelerated as
efficiently as electrons to a maximum energy determined by the size
and magnetic field of the emitting region, we calculate the emissivity
of neutrinos produced by protons that interact with the external
radiation field intercepted by the jet. The external radiation field
provides the most important target photons for photomeson production
of high-energy neutrinos in flat spectrum radio quasars (FSRQs).
Because of photomeson interactions with this field, km$^2$ neutrino
telescopes are predicted to detect $\gtrsim 0.1$-1 neutrinos per year
from blazars such as 3C 279. BL Lac objects are weaker neutrino
sources if, as widely thought, their $\gamma$-ray emission is due to
Compton-scattered synchrotron (SSC) radiation.
\end{abstract}

\section{Introduction}

Blazars are among the most powerful accelerators of relativistic
particles in nature, as shown by EGRET observations of 100 MeV - GeV
emission from over 60 FSRQs and BL Lac objects. Consequently, blazars
represent a potential source of high-energy neutrinos (see Gaisser et
al.\ 1995 and Halzen 2001 for reviews). The bright, highly variable
$\gamma$ radiation from blazars argues for the existence of emission
regions with high photon energy densities and intense nonthermal
particle populations, which are needed for efficient photomeson
production (e.g., Mannheim and Biermann 1992; Mannheim et al.\
2001). Previous papers have considered internal synchrotron photons as
targets for high-energy proton interactions (e.g., Mannheim
1993). Here we show that UV photons from external radiation fields,
which have earlier been proposed as a target photon source for
nonthermal electrons to produce Compton-scattered gamma-rays (e.g.,
Dermer and Schlickeiser 1993; Sikora et al.\ 1994), also provide the most
important photon source for photomeson production of $\gtrsim 30$ TeV
neutrinos in FSRQs.

We use the measured spectral fluxes and variability time scales of
synchrotron and Compton radiation from the blazar 3C 279 to determine
physical parameters of its gamma-ray emission region. Lower limits to
the Doppler factor $\delta$ are defined by the condition that the
emitting region be transparent to gamma rays. Equipartition arguments
and observed fluxes of the synchrotron and Compton components are used
to determine the magnetic field strength $B$ in the plasma blob.
Assuming efficient acceleration of power-law energetic protons to a
maximum energy limited by the magnetic field and size scale of the
region, we calculate spectral emissivities of pion decay neutrinos
formed through photomeson production. Because the properties of the
external radiation field are independent of the bulk Lorentz factor
$\Gamma$ of the radiating region, neutrino emissivity calculations
using duty cycle estimates give accurate neutrino flux estimates
during low $\gamma$-ray states. We predict detectable fluxes of
neutrinos from $\gamma$-ray loud FSRQs using km$^2$ neutrino telescope
arrays such as IceCube.

\section{Constraints on Source Parameters}

\subsection{Synchrotron and external photon energy densities}

The comoving synchrotron photon energy density is given by $u^\prime_s
\cong$ $ L_s/$$(2\pi r_b^2 c \delta^4)$, where $L_s$ is the observed
bolometric synchrotron luminosity, $r_b$ is the comoving radius of the
blob, here assumed spherical, and primes denote comoving
quantities. The ratio of the SSC and synchrotron luminosities
$L_{SSC}/L_s \approx u^\prime_s/u_B$, where $u_B = B^2/8\pi$.  For
simplicity, we consider the regime where the quasi-isotropic scattered
external radiation field dominates the direct disk radiation field
(Dermer and Schlickeiser 1994).

If the $\gamma$-rays are due to external Compton-scattered radiation,
then $L_{EC}/L_s \cong \delta^2 u_{ext}/u_{B}\cong
(\delta/\Gamma)^2u^\prime_{ext}/u_B$ (Sikora 1997; Dermer, Sturner,
and Schlickeiser 1997). Combining these two expressions gives
$u^\prime_{ext}/u^\prime_s $$\cong$ $ (\Gamma/\delta)^2$ $
(L_C/L_{SSC})$ $ \equiv a (\Gamma/\delta)^2 \cong a$ when viewing
within the beam of the jet ($\theta \lesssim \Gamma^{-1}$).
Consequently $u^\prime_{ext} \cong$ $ a L_s$ $/(2\pi r_b^2c
\delta^4)$, where $a \equiv L_{EC}/L_{SSC}$ The energy of the external
photons in the comoving frame is $\e_{ext}^\prime \cong
\Gamma\e_{ext}$. Models taking into account SSC and external Compton
(EC) components show that a complete spectral fit requires
synchrotron, SSC and EC components (B\"ottcher 1999; Hartman et
al. 2001). Thus $a \sim 0.1$-1 for BL Lac objects and $a \sim 1$-10
for FSRQs.

The spectral energy density $u_{\rm ph}(\epsilon^\prime)\equiv m_e c^2
\epsilon^{\prime^2} n^\prime_{\rm ph}(\epsilon^\prime)$ of the photons
in the blob frame is given in terms of the $\nu F_\nu$ flux
$f(\epsilon)$ through the relation
\begin{equation}
u_{\rm ph}^{\prime}(\epsilon^\prime)\cong 
\frac{2 d_{\rm L}^2  f(\epsilon)}
{r_{\rm b}^2 \, c \, \delta^4}\;\cong 
\frac{2 d_{\rm L}^2  (1+z)^2 f(\epsilon)}
{ c^3 \, t_{var}^2({\rm d})\delta^6}\; .
\label{eq1}
\end{equation}
where we relate the measured variability time scale $t_{var}$ to the
blob radius $r_b$ through the expression $r_b \simeq ct_{var}\delta$
$/(1+z)$, $t_{var}({\rm d})$ is the observed variability time scale in
days, and $z$ is the redshift. The expression $\epsilon^\prime =
\epsilon (1+z)/\delta$ relates the photon energies measured in the
source and observer frames.

\subsection{Magnetic field estimates}

The magnetic field in the blob can be estimated by noting that the
total soft photon energy density $u_{ph}^\prime = u^\prime_s +
u^\prime_{ext} \simeq u_B(L_C/L_s)$, where $L_C = L_{EC}+L_{SSC}$ is
the total Compton power. Hence $u_B = (L_s/L_C)u^\prime_{ph} =
(L_s/L_C)(1+ a)u^\prime_{s}$. Using the earlier expressions for
$u^\prime_{s}$ and $r_b$ gives
\begin{equation}
B\simeq {2(1+z)\sqrt{1+a}\over c\delta^3 t_{var}}\;({L_s\over
c})^{1/2}\;({L_s\over L_C})^{1/2}\;.
\label{B}
\end{equation}

An independent estimate for $B$ can also be made by introducing the
equipartition parameter $\eta = u^\prime_{\rm el}/u_{\rm B}$ for the
ratio of nonthermal electron to magnetic energy densities in the jet.
Thus $\eta \sim 1$ corresponds to equipartition, though with electrons
only. For the measured synchrotron flux density $S_{\nu} \propto
\nu^{-\alpha_{\rm r}}$ with $\alpha_{\rm r}\simeq 0.5$ 
at $\nu \lesssim 10^{13}\,\rm Hz$, this gives 
\begin{equation}
B \cong 220 \; [{\ln({\gamma_{max}\over \gamma_{min}})\over
10}]^{2/7}\;{\nu_{12}^{1/7} d_{28}^{4/7}[S({\rm
Jy})]^{2/7}(1+z)^{5/7}\over \eta^{2/7}[t_{\rm var}({\rm
d})]^{6/7}\delta^{13/7}}\;
\label{Beq}
\end{equation}
in Gauss, where $S({\rm Jy})$ is the flux density measured at $10^{12}
\nu_{12}$ Hz.

\subsection{$\gamma\gamma$ transparency constraints on Doppler factor}

Transparency to $\gamma\gamma$ pair-production attenuation requires
that $\tau_{\gamma\gamma}(\epsilon^\prime) \cong 2\sigma_{\rm
T}n^\prime_{ph}(2/\epsilon^\prime)r_b/(3\epsilon^\prime) < 1$, which
can be rewritten as a lower limit on $\delta$ using equation
(\ref{eq1}).

\subsection{Photopion and neutrino production}

Energy losses of relativistic protons (and neutrons) are
calculated on the basis of standard expressions
(see, e.g., Berezinskii and Grigoreva 1988) for the 
cooling time of relativistic protons due to photopion production
in $p \gamma$ collisions. If the ambient photons have spectral density 
$n^\prime_{\rm ph}(\ep) $, then
$$
t_{\rm p \gamma}^{-1}(\gamma_p) = \int_{\frac{\epsilon_{\rm th}}
{2 \gamma_p}}^{\infty}{\rm d}\ep\;{ 
\frac{c \, n^\prime_{\rm ph}(\ep)}{2 \gamma_p^2 \epsilon^{\prime 2} }
\int_{\epsilon_{\rm th}}^{2\ep \gamma_p}{\rm d} \epsilon_{\rm r}\;{
\sigma(\epsilon_{\rm r}) K_{\rm p \gamma}(\epsilon_{\rm r}) \epsilon_{\rm r} 
  }\, }
\, , 
$$ where $\gamma_p$ is the proton Lorentz factor, $\epsilon_{\rm r}$
is the photon energy in the proton rest frame, $\sigma(\epsilon_{\rm
r})$ is the photopion production cross-section, $\epsilon_{\rm
th}\approx 150 \,\rm MeV$ is the threshold energy for the parent
photon in the proton rest frame, and $ K_{\rm p \gamma}(\epsilon_{\rm
r})$ is the inelasticity of the interaction.  The latter increases
from $ K_{1}\approx 0.2 $ at energies not very far above threshold
where the single pion production channels dominate, to $K_{2}\sim
0.5-0.6 $ at larger values of $\epsilon_{\rm r}$ where multi-pion
production dominates (e.g. see Berezinskii and Grigoreva 1988, M\"ucke
et al.\ 1999).

A detailed recent study of this photohadronic process has been given
by M\"ucke et al. (1999). In order to simplify calculations, we
approximate the cross-section $\sigma(\epsilon_{\rm r})$ as a sum of 2
step-functions $\sigma_{1}(\epsilon_{\rm r})$ and
$\sigma_{2}(\epsilon_{\rm r})$ for the (total) single-pion and
multi-pion channels, respectively, with $\sigma_{1}=380~ \mu \rm b$
for $ 200 \,\rm MeV \leq \epsilon_{\rm r}
\leq 500 \, MeV$ and $\sigma_1 = 0$ outside this region, whereas
$ \sigma_2 = 120 ~\mu \rm b$ at $\epsilon_{\rm r} \geq 500 \,\rm MeV$.
The inelasticity is approximated as $K_{\rm p\gamma} =K_1$ and $K_{\rm
p\gamma}=K_2$ below and above 500 MeV.  The results of our
calculations in this simplified approach are found to give good
agreement with earlier calculations (Berezinskii and Grigoreva
1988). In particular, the time scales for photopion interactions of
ultra-relativistic cosmic rays with the CMBR are accurately reproduced
(see also Stanev et al.\ 2000). This approach also works well for a
broad power-law distribution of field photons $n^\prime_{\rm ph}
\propto \epsilon^{\prime-\alpha_{\gamma}}$ for different photon
spectral indices $\alpha_\gamma$, and readily explains the significant
increase in the mean inelasticity of incident protons (or neutrons)
from $\langle K_{\rm p\gamma}\rangle \simeq 0.2 $ for steep photon
spectra $n_{\rm ph}^\prime(\ep )$ with $ \alpha_{\gamma} \geq 1 $, to
$\langle K_{\rm p\gamma}\rangle \rightarrow 0.6 $ for hard spectra
with $ \alpha_{\gamma} \leq 1 $ (M\"ucke et al.\ 1999).

Calculations of the spectra of secondary particles and photons ($\nu,
\; \gamma, \; e$) are done in the $\delta$-function approximation,
assuming that the probabilities for producing pions of different
charges ($\pi^0$, $\pi^{+}$ and $\pi^{-}$) are equal for the
multi-pion interaction channel. To correctly apply the
$\delta$-function approximation, one has to properly take into account
the different inelasticities of the multi-pion and single-pion
production channels, which could explain the statement of M\"ucke et
al.\ (1999) that the $\delta$-function approximation does not work
well in the case of hard photon spectra.

\subsection{Photon fields in the gamma-ray production region}  

We consider 3C 279 as a prototype FSRQ.  The broad-band radiation
spectra produced by the relativistic ejecta in the jet of 3C 279 is
generally thought to consist of a low energy synchrotron component and
high energy Compton component. In 3C 279, the synchrotron component
dominates at observed frequencies $\nu \ll 10^{16} \,\rm Hz$. We
assume that the $\nu \gg 10^{11}$ Hz synchrotron emission is
cospatially located with the gamma-ray production region. This
assumption is supported by observations of correlated variability
between the synchrotron and Compton components in some BL Lac objects
and FSRQs. The high-energy Compton emission consists of an SSC
component and an external Compton (EC) component due to target photons
that originate from outside the jet; for example from the disk, torus,
scattered-disk radiation, reflected synchrotron radiation, etc.

For calculations of $n^\prime_{\rm ph}(\ep )$ from 3C 279, we
approximate the differential photon flux $\Phi(\epsilon)$ observed
during the powerful flare of 1996 (Wehrle et al.\ 1998) in the form of
a continuous broken power-law function, with power-law photon indices
$\alpha_{1}\cong 1.5$, $\alpha_{2}\cong 2.45$, and $\alpha_{3}\cong
1.6$ at frequencies $\nu \leq \nu_1=10^{13}\,\rm Hz$, $\nu_1 \leq \nu
\leq \nu_2 =10^{16}\,\rm Hz$, and $\nu \geq \nu_2$, respectively.  The
$\nu F_\nu$ energy flux $f(\epsilon) \equiv
\epsilon^2 \, \Phi(\epsilon)$ of the synchrotron radiation 
reaches a maximum value of $1.66\times 10^{-10} \,\rm erg\, cm^{-2}
s^{-1}$ at $\nu = \nu_{1}$. The maximum flux of the Compton component
is $\sim 10$ times larger than the peak synchrotron flux during the
flare, and is reached at $\sim 500 \,\rm MeV$. We suppose that the SSC
component makes a substantial contribution to the hard X-ray emission
in 3C 279 (Hartman et al.\ 2001), and therefore that $a \cong 10$.
   
The redshift of 3C 279 is $z = 0.538$, and its luminosity distance is
$d_{\rm L}(z)\cong 1.05\times10^{28}\,\rm cm$ for an $\Omega_m = 0.7$,
$\Omega_\Lambda = 0.3$ cosmology with a Hubble constant of 65 km
s$^{-1}$ Mpc$^{-1}$. The measured $> 10^{16}$ Hz flux during the 1996
flare implies that $\delta \gtrsim 4.9 [E_{ph}({\rm
GeV})^{0.117}/[t_{var}({\rm d})]^{0.19}$ from the $\gamma\gamma$
transparency arguments. Thus $\delta \approx 5$ corresponds to the
minimum Doppler-factor required for relativistic blobs in 3C 279 to
become transparent to absorption of the observed GeV photons in the
radiation field inside the blob.

Expressing $r_b$ via the observed variability timescale $t_{\rm
var}\sim$ 1-2 days as observed by EGRET (Wehrle et al.\ 1998), one
finds from equation (\ref{eq1}) that
\begin{equation}
u_{\rm ph}^{\prime}\sim 16 f_{-10} \, 
[t_{\rm var}({\rm d})]^{-2} \, \delta_{5}^{-6}
\; {\rm erg~cm}^{-3} \, .
 \end{equation} Here $f_{-10} \equiv$ $ f(\epsilon)/(10^{-10}$erg
cm$^{-2}$ s$^{-1}$), and $\delta_{5}\equiv \delta / 5$.  From the
observed flux of synchrotron radiation from 3C 279 with photon number
index $\alpha = 1.5 $ at $\nu \ll 10^{13}
\,\rm Hz$, we deduce $ B(\rm G) \simeq 20 \;\eta^{-2/7} [t_{\rm var}
(\rm d)]^{-6/7} \, 
\delta_{5}^{-13/7}$
from equipartition arguments. The estimation of $B$ from equation
(\ref{B}) gives $B(\rm G) \cong 32\; t^{-1}_{var}(\rm
d)\delta_{5}^{-3}$, using $L_s = 3\times 10^{47}$ ergs s$^{-1}$,
$L_C/L_s \cong 10$, and $a \cong 10$.

The external radiation energy density in the blob rest frame is, using the 
expression for $B$ derived from equipartition arguments, given by 
\begin{equation} 
u_{ext}^{\prime}\simeq 20 \, ({L_{EC}\over L_s}) \, \eta^{-4/7} 
[t_{\rm var}(\rm d)]^{-12/7}\delta_{5}^{-26/7} \,{\rm erg~cm}^{-3} \, . 
\end{equation}
For the calculations we assume $\eta=1$ and take  $a =  10$.
 
\section{ Calculations and Results}

We assume that the spectrum of the external UV radiation field arises
from a Shakura/Sunyaev (1973) optically-thick accretion disk model
that is scattered by broad-line region clouds.  This disk model has
flux density $F(\epsilon ) \propto
\epsilon^{1/3}$ up to a maximum photon energy $\e_{max}$ 
determined by the innermost radius of the blackbody disk and
properties of the central engine. We take $m_ec^2 \e_{max} = 35$ eV
(Dermer and Schlickeiser 1993).  The energy density of the external
radiation in the comoving plasma blob frame can be determined by
noting that $f(\epsilon \gtrsim 1)/f(\epsilon\lesssim 1)$ corresponds
to the ratio $L_{EC}/L_s$ of the directional external Compton and
synchrotron powers as measured along the jet axis.

\begin{figure}[t]
\vspace*{2.0mm} 
\includegraphics[width=8.3cm]{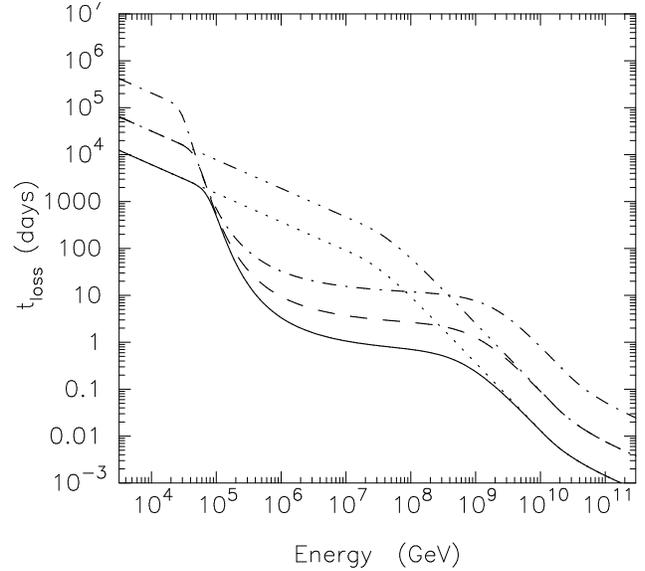} 
\caption{
Photomeson interaction energy loss time scale of protons, calculated
for spectral fluxes observed from 3C~279 (see text) and $t_{\rm
var}=1\,\rm d$, assuming 3 different Doppler-factors for the jet:
$\delta = 7$ (solid curve), $\delta = 10$ (dashed curve), and $\delta
= 15$ (dot-dashed curve). The dotted and triple-dot--dashed curves are
calculated for $\delta = 7$ and $\delta = 10$, respectively, when $p
\gamma$ interactions with the synchrotron radiation field alone are
considered.}
\end{figure}

Fig.\ 1 shows energy loss timescales (in the observer frame) of
protons due to photopion production in a jet of 3C~279 calculated for
3 different Doppler-factors: $\delta = 7$ (solid curve), $\delta = 10$
(dashed curve), and $\delta = 15$ (dot-dashed curve).  The dotted and
triple-dot--dashed curves show the photopion timescales corresponding
to interactions with the synchrotron radiation only, for the cases
$\delta =7$ and $\delta = 10$.

\begin{figure}[t]
\vspace*{2.0mm} 
\includegraphics[width=8.3cm]{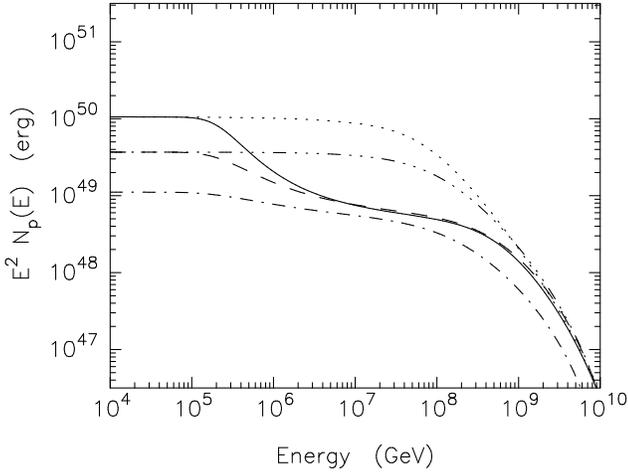} 
\caption{
The energy distribution of relativistic protons $N_{\rm p}$ calculated
for the same jet Lorentz-factors $\delta = 7 $, 10, and 15 and
assumptions for the external and internal radiation fields as made in
Fig.\ 1, assuming a power-law injection of relativistic protons with
number index $\alpha_{\rm p}=2$ during $\Delta t = 2 \,$ weeks with a
power $L_{\rm p}=10^{48} \delta^{-4}$ ergs s$^{-1}$.  }
\end{figure}

Fig.\ 2 shows the energy distributions of relativistic protons $N_{\rm
p}$ in the ejecta/blobs (for the same cases as in Fig.\ 1), which are
calculated assuming power-law injection of relativistic protons with
number index $\alpha_{\rm p}=2$ on observed timescales $\Delta t = 2
\,$ weeks. The total injection power of the protons $L_{\rm
p}=10^{48}\delta^{-4}$ ergs s$^{-1}$.  Note that $10^{48}\,\rm ergs$
s$^{-1}$ is the characteristic {\it apparent} luminosity of the flares
detected by EGRET from 3C~279. In calculations of $N_{\rm p}$ we take
into account the photohadron interaction energy losses, as well as the
escape losses of the protons in the Bohm diffusion limit.  This limit
is more restrictive than the condition that the gyroradius of the
highest energies of accelerated protons should not exceed the blob
size (Hillas 1984).
  
Fig.\ 3 shows the expected energy fluxes of the neutrinos produced by
protons in Fig.\ 2. For the fluxes in Fig.\ 3, the total number of
neutrinos that could be detected by a $1 \,\rm km^2$ detector during 2
weeks of the flare, using neutrino detection efficiencies given by
\citep{ghs95} is 0.19, 0.12 and 0.05 for the cases of $\delta=7$, 10
and 15, respectively.  It is important to note that if the external UV
field is neglected, these numbers would be significantly reduced. We
calculate the number of neutrinos that would be detected by a km$^2$
array ranging from $1.7\times 10^{-2}$ to $3\times 10^{-3}$ for the
fluxes in Fig.\ 3, which would not leave a realistic prospect for the
detection of at least 2-3 neutrinos, which is required for a positive
detection of a source. BL Lac objects, which have weak broad line
regions and, therefore, a weak scattered external radiation field,
should consequently be much weaker neutrino sources.

Because the properties of the external radiation field are insensitive
to the value of the blob Lorentz factor, neutrino production in the
quiescent state can be accurately estimated on the basis of properties
of the external radiation field derived during the flaring state.  For
a typical duty factor of the flares $\sim $ 1-2 months per year, and
considering the additional neutrino production during the quiescent
phase, we can reasonably expect that IceCube or other km$^2$ array may
detect several neutrinos per year from 3C 279-type blazar jets with
$\delta \lesssim 10$.  Allowing for the possibility that the
nonthermal protons may have an overall power larger than that for
primary electrons improves the prospects for neutrino detection.

It is important to notice that the external photon field reduces very
significantly the characteristic energy of protons for effective
photopion production, compared with the case of the internal
synchrotron radiation field. In fact, acceleration of protons only to
$E_{\rm p}\sim 10^{15}-10^{16}\,\rm eV$ is required for efficient neutrino
production through photomeson interactions, as can be seen from Fig.1.
This effect permits us to predict that a km$^2$ array will detect
high-energy neutrinos from FSRQs without requiring acceleration of
protons to ultra-high energies $E_{\rm p}\gg 10^{18}\,\rm eV$.

\begin{figure}[t]
\vspace*{2.0mm} 
\includegraphics[width=8.3cm]{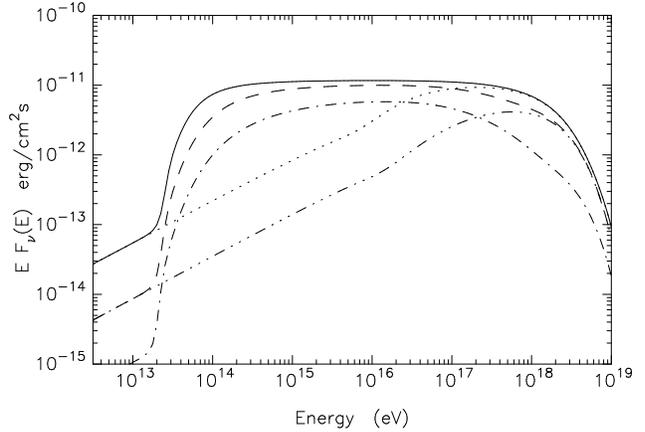} 
\caption{The fluxes of neutrinos expected due to photomeson interactions 
of protons shown in Fig.\ 2 with the external (UV) and internal
(synchrotron) photons in the jet of 3C~279 for Dopper factors
$\delta =7$ (solid curve) $\delta = 10 $ (dashed curve), and 
$\delta =15 $ (dot-dashed curve). The dotted and triple-dot--dashed 
curves show the calculated neutrino fluxes if the external
photon field in the jet is neglected.}
\end{figure}

\end{document}